# The use of cluster quality for track fitting in the CSC detector

Erez Etzion, David Primor, Giora Mikenberg, Nir Amram and Hagit Messer, *Fellow, IEEE*

*Abstract*— The new particle accelerators and its experiments create a challenging data processing environment, characterized by large amount of data where only small portion of it carry the expected new scientific information. Modern detectors, such as the Cathode Strip Chamber (CSC), achieve high accuracy of coordinate measurements (between 50 to 70 microns). However, heavy physical backgrounds can decrease the accuracy significantly. In the presence of such background, the charge induced over adjacent CSC strips (cluster) is different from the ideal Matheison distribution. The traditional least squares method which takes the same ideal position error for all clusters loses its optimal properties on contaminated data. A new technique that calculates the cluster quality and uses it to improve the track fitting results is suggested. The algorithm is applied on test beam data, and its performance is compared to other fitting methods. It is shown that the suggested algorithm improves the fitting performance significantly.

## I. INTRODUCTION

The LHC, the largest hadron collider ever built, presents new challenges for physicists and engineers. With the anticipated luminosity of the LHC, one expects as many as one billion total collisions per second, of which at most 10 to 100 per second might be of potential scientific interest. The track reconstruction algorithms applied at the LHC will therefore have to reliably reconstruct tracks of interest in the presence of background hits.

One of the two major, general-purpose experiments at LHC is called ATLAS. Since muons are among the most important particles to be detected as a sign of new phenomena, a stand-alone muon spectrometer system is being built for ATLAS [1]. The ATLAS muon spectrometer is located in high radiation background environment which makes the muon tracking a very challenging task. In the inner station of the ATLAS muon spectrometer, Cathode Strip Chambers (CSC) system is employed for the precision measurement of muon tracks in the forward region ($2.0 < \eta < 2.7$ where $\eta$, the pseudo-rapidity angle, is defined as $\eta = -\ln tag(\vartheta/2)$ and $\vartheta$ is the angle between the muon and the beam direction). In this region the fluxes of photons and neutrons can reach 2800Hz/cm² which will cause uninteresting contaminating hits to appear close to the muon tracks.

Though many particle tracking algorithms were developed during the years by the High Energy Physics (HEP) community (see for example [2]), the requirement of high accuracy in an environment with such a high background level is still very demanding. In the presence of such background, the charge induced over adjacent strips (cluster) of the CSC is different from the ideal distribution. Thus, the cluster position error can no longer assumed to be as an error of an ideal cluster. Track fitting that uses the traditional least squares (LS) method takes into account the same ideal position error for all clusters, and thus it loses its optimal properties on contaminated data. We suggest a new track fitting technique, which calculates the cluster quality and distinguish between "clean" and "dirty" clusters. It is shown that fitting algorithm that uses only the "clean" clusters, results in better performance than other methods such as traditional least squares (LS) and even weighted least squares (WLS), and robust fitting. The rest of the paper is organized as follows: on section II the CSC detector is described as well as the experimental set up used to evaluate the fitting algorithm. On section III the method of calculating the cluster quality is described. On section IV the different track fitting techniques are described and compared making use of test beam data. The results are discussed on section V.

## II. THE CSC DETECTOR AND THE TEST BEAM

### A. The ATLAS CSC

The ATLAS CSC detector [1] consists of four-layer chambers that provide a position measurement based on charge-interpolation. When an energetic particle, presumably a muon, passes through the chamber, it ionizes a local region of the gas that fills the chamber. The ionized cluster of electrons drifts toward a nearby anode wire and a charge avalanche is established. The charge avalanche induces charge on two sets of cathode strips that are mutually perpendicular (there are 192 strips oriented orthogonal to the anode wires and 48 strips parallel to it). The induced charge is spread out over several adjacent strips; each strip receiving a fraction of the total induced charge. The spread of strips that receive charge is called a hit-cluster. The concept is that with the knowledge of the interpolated total charge passing through a layer, calculating the relative magnitudes of both the charge on each strip and the position of the strip in the hit-cluster will provide the information required to find the centroid of the charge. The centroid is the position at the chamber where the ionization cluster originated, thus, the position of the energetic particle's track. Figure 1 describes the charge induction over the precision strips.

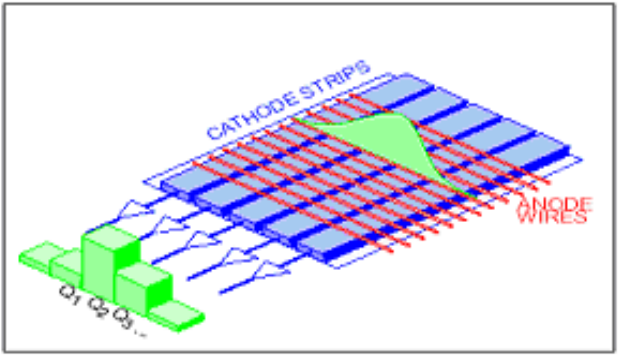

Fig. 1. Charge distribution over the precision strips [1]

*B. The Test Beam*

The background environment of the LHC experiments is expected to be very high, especially in the forward regions. In order to study the performance in the presence of high backgrounds, a CSC chamber was tested at the GIF radiation facility at CERN during July 2004. This facility provides a high energy muon beam in conjunction with a variable high intensity radioactive gamma source $664\,GBq\,^{137}Cs$. Variable lead absorbers were used to change the radiation intensity. The beam illuminated area of the chambers was determined by trigger counters. A detailed description of the summer 1998 test beam setup for the CSC prototype is described in [3]. The main difference between that test and more recent ones is the lack of Si telescope in the second experiment.

III. A QUALITY OF A CSC CLUSTER

While most of the commonly used algorithms for track finding and fitting assume a well understood shape of the charged clusters, even in presence of a high radiation background, the present philosophy has been to use all available information, combined with the inherent high efficiency of the CSC detectors, for track finding [4], while making use of only clean charged clusters for track fitting. It will be shown that such a procedure provides a more accurate measurement of tracks, since the "dirty" clusters have carry large systematic errors in their description.

Since minor disturbance of the signal on at least one of the cluster strips may disturb the measured position, we classify the hit clusters into "clean" and "dirty" clusters. The "clean" clusters are those which are well separated from other clusters, and have the ideal charge distribution as expected according to the Matheison distribution [5] described in Figure 2. The "dirty" clusters are those which are either close to other clusters, or have a spatial charge different from the Matheison distribution.

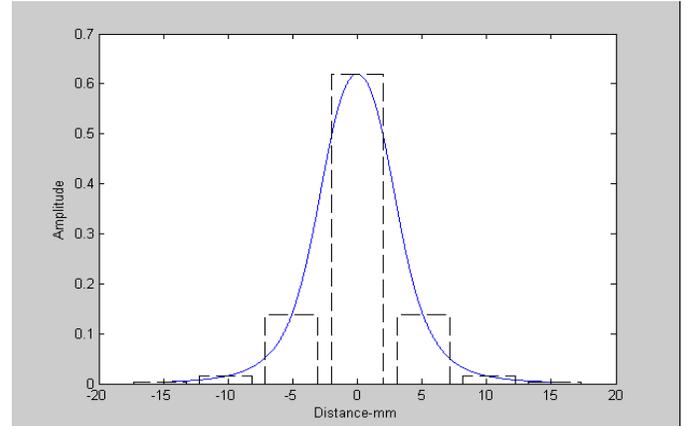

Fig. 2. The Matheison spatial distribution (line) and an example of the charge induced over the CSC strips according to this distribution (dashed line)

It can be shown that charge induced over $N$ strips by a particle traveling perpendicularly through the detector can be modeled as:

$$y(n) = AS_x(n; x_p) + r(n) \qquad (1)$$

where n is the spatial sample (the strip) and $S_x(\square)$ is the spatial signal shape over the detector strips which follows the Matheison distribution. $r(n)$ is a general expression for noise and interferences. $A$, $x_p$ - the amplitude, and the hit position - are the parameters to be estimated. In the case of a "clean" cluster, $r(n)$ is a vector of Gaussian white noise without other interferences. In this case, the probability of the data $\mathbf{Y} = [y(0), y(1),..y(N)]$ is:

$$P_\mathbf{y}(x_p, A) = \prod_{i=1}^{N} \frac{1}{(2\pi\sigma^2)} \exp[-\frac{1}{2\sigma^2}(\mathbf{Y}[i] - \mathbf{C}(x_p)[i]A)^2] =$$
$$(2\pi\sigma^2)^{-N} \exp[-\frac{1}{2\sigma^2}\sum_{i=1}^{N}(\mathbf{Y}[i] - \mathbf{C}(x_p)[i]A)^2] = \qquad (2)$$
$$(2\pi\sigma^2)^{-N} \exp[-\frac{1}{2\sigma^2}|\mathbf{Y} - \mathbf{C}(x_p)A|^2]$$

where $\mathbf{C}(x_p) = [S(0-x_p), S(1-x_p),...,S(N-1-x_p)]^T$ and $\sigma^2$ is the noise variance. The parameters $A$ and $x_p$ are estimated

using the maximum likelihood (ML) method:

$$(\hat{A}, \hat{x}_p) = \arg\min_{A, x_p} |Y - C(x_p)A|^2 \quad (3)$$

and its solution is given by:

$$\hat{A} = \frac{Y^T C(x_p)}{C(x_p)^T C(x_p)} \quad (4)$$

Substituting (4) into (3) shows that the maximum likelihood estimation of $x_p$ is obtained by solving the following one dimensional optimization problem:

$$\hat{x}_p = \arg\max_{x_p} \frac{(Y^T C(x_p))^2}{C(x_p)^T C(x_p)} \quad (5)$$

The quality of the cluster is determined using a one-sided test; that is, the cluster is considered as a "clean" cluster if it is well separated from other clusters and:

$$P_y(\hat{x}_p, \hat{A}) > \lambda \quad (6)$$

where $\lambda$ is a predefined threshold. Practically we use the quality value $Q$ as:

$$Q \equiv \arg\max_{x_p} \frac{(Y^T C(x_p))^2}{C(x_p)^T C(x_p)} > \lambda' \quad (7)$$

where $\lambda'$ is a predefined threshold.

In order to decide about the range of the threshold $\lambda'$, we calculate $Q$ of (7) for two scenarios using the test beam data. For the first scenario we use data received when the gamma source was not activated (clean data) and in the second scenario we use the data received when the source was activated (noisy data). In Figure 3 we plot the quality values for the two scenarios. It can be seen that a threshold value $\lambda'$ in the range between 30 and 40 might be a reasonable decision.

It is shown in [6] that the position estimation error for "clean" and "dirty" clusters are both Gaussian distributed. While the standard deviation (STD) error of the "clean" clusters is about 70 microns, the STD error of the "dirty" clusters is much larger. Since track fitting with least squares (LS) method is not optimal for hits with different errors, the use of cluster quality has a potential for track fitting improvement. In the next session we will compare several track fitting techniques with and without the quality information.

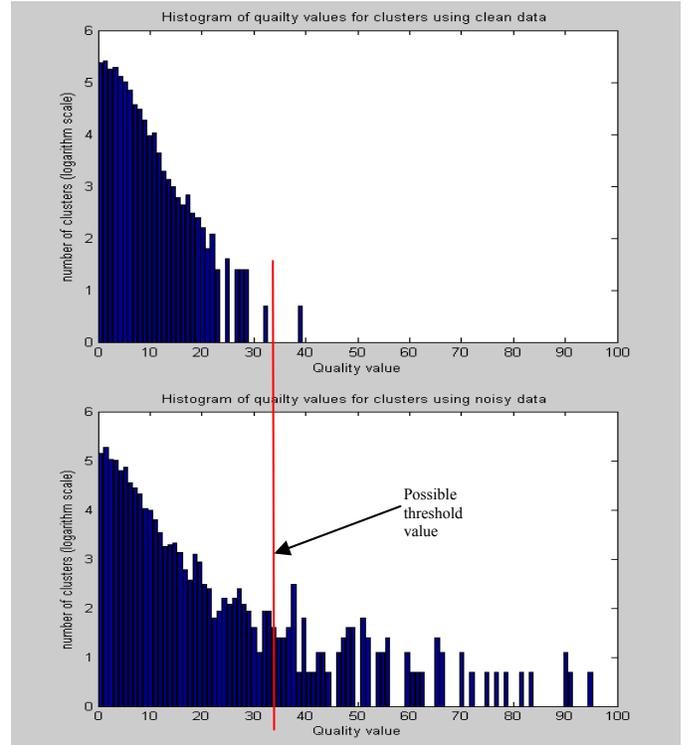

Fig. 3. Histogram of the quality values for two scenarios. The top histogram is calculated using "clean" data from the test beam, while the bottom histogram is calculated using "dirty" data. In order to emphasize the difference in the tails, we used a logarithmic scale. The tail in the bottom histogram is longer (till about 1000), and the maximum value of 100 is picked for display reasons.

## IV. TRACK FITTING

### A. The track fitting techniques

In order to evaluate the benefit of using the segment quality, we compare different track fitting techniques realized on the test beam data. The first two techniques listed below do not utilize the cluster quality while the rest do use it:

*Least squares (LS) fitting* – in this case all clusters were used for the track fitting. The position of each cluster was calculated using the ratio algorithm [3]. Then, the cluster positions were taken for a least squares fit.

*Iterative LS fitting* – this iterative algorithm was suggested in [3]. A linear regression is applied for all clusters as a first step. Then, the cluster with the biggest residual is omitted, and the linear regression is applied again. Since the CSC has only four layers, the regression procedure can be applied maximum three times.

*Weighted Least squares (WLS) fitting* – in this case the cluster quality was calculated, and different weights were given to each cluster. Where the "clean" cluster got weight equal to one, the "dirty" clusters got weight equal to 0.1.

*Robust fitting* - Following the mathematical inference described in [7], the iterative least-squares equations using the Tukey's bi-weights [8] are:

$$w_i = \begin{cases} (1-(u_i)^2)^2 & |u_i|<1 \\ 0 & |u_i|>1 \end{cases} \quad (7)$$

where $u_i = \dfrac{\varepsilon_i}{\sigma \cdot S}$, $\varepsilon_i$ is the residual between the cluster and the fitted track for the i$^{th}$ iteration, $\sigma$ is the robust standard deviation and $S$ is the cutoff or tuning constant. The WLS fitting is applied for initial conditions.

*Restricted LS fitting* - only the "clean" clusters are used for the track fitting process. In the cases where there are no "clean" clusters, all clusters are used.

### B. Comparison of the fitting methods

In order to compare the different techniques we used the test beam data with an average background rate of 3 KHz/cm2 (about five times larger than the ATLAS expected background rate). In principle the algorithm results should have been compared to the real tracks ("truth"). However since in the discussed test bean there was no reference measurement to cross check the information collected with the CSC, the "truth" information was not available. The test beam experiment produced many data files with different background conditions. Therefore, the data collected without active gamma-ray source (referred as "clean data") was used as our "truth" reference; it was combined with data taken when the source was activated ("noisy data"). Only events with four hits, one from each layer of the CSC were taken as valid "truth" data points. Then, the position of each cluster was calculated using the ratio algorithm [3] and a linear regression was applied for track finding. This track was taken to be the "truth". We define a track fitting efficiency as the ratio $n_{CSC}^{(5\sigma)}/N$, where $N$ denotes the total number of tracks and $n_{CSC}^{(5\sigma)}$ is the number of reconstructed tracks which is consistent with the track within five times the track resolution. The track resolution is the standard deviation error of the estimated muon tracks position for clean events. Figure 4 describes the track fitting efficiency for the different methods. It can be seen that the *Restricted LS* method gets the best track fitting efficiency. This method requires calculating the cluster quality and uses only the "clean" clusters. If there is only one "clean" cluster, the distance between the cluster position and "truth" is used for the efficiency calculation. Although the *WLS* and *Robust fitting* algorithms use all the clusters with appropriate weights, they have smaller efficiency compared to the *Restricted LS* fitting algorithms. This can lead to the conclusion that the "dirty" clusters should not be used in the line fitting process. The *Iterative LS* method applies the *LS* calculation three times for each line. Its efficiency is low compared to the techniques which use the cluster quality. It can be seen that the *LS* technique has a significantly low efficiency compared to the other fitting methods.

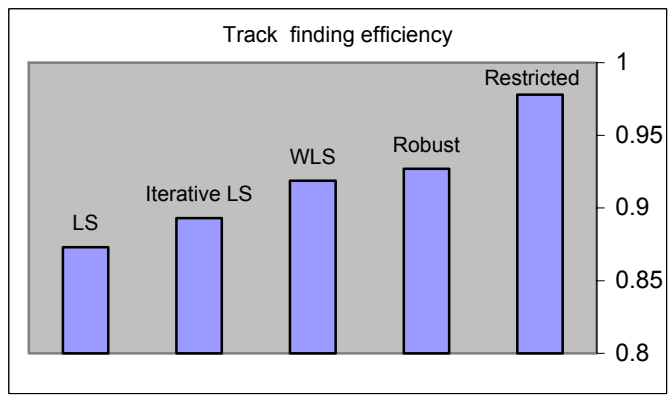

Fig. 4. The track fitting efficiencies for the different methods.

## V. DISCUSSION

The local track identification is one of many new challenges in experimental HEP. Several fitting methods are compared for estimating the muon track in the ATLAS CSC detector. The classification of the cluster into "clean" and "dirty" was proved to be a very useful technique for improving the track fitting efficiency. Unfortunately, at high background rate, we can get less than two "clean" clusters, a situation which leads to a poor local track fitting. In these cases the use of information from other components of the ATLAS detector can be of great help, and an overall fitting that uses only the "clean" clusters might be the right technique.

### ACKNOWLEDGMENT

We would like to thank the CSC community, and especially to Prof. Venetios Polychronakos from Brookhaven National Lab for giving us the access for the CSC test beam data and many good advises for a better understanding of the CSC detector.